\documentclass[ 
reprint,
superscriptaddress,
 amsmath,amssymb,
prb,
]{revtex4-2}

\usepackage{graphicx}
\usepackage{dcolumn}
\usepackage{bm}
\usepackage[final]{changes} 
\definecolor{Changes@Color}{named}{red} 
\usepackage{xcolor}
\usepackage[normalem]{ulem}

\usepackage[utf8]{inputenc}   
\DeclareUnicodeCharacter{2212}{-}

\begin{document}

\preprint{APS/123-QED}

\title{Quantifying Trapped Magnetic Vortex Losses in Niobium\\ Resonators at mK Temperatures}

\author{D. Bafia}
 \altaffiliation{dbafia@fnal.gov}
\affiliation{Fermi National Accelerator Laboratory, Batavia, Illinois 60510, USA}
\author{B. Abdisatarov}
\affiliation{Fermi National Accelerator Laboratory, Batavia, Illinois 60510, USA}
\affiliation{Department of Electrical and Computer Engineering, Old Dominion University, Norfolk, Virginia 23529,
USA}
\author{R. Pilipenko}
\affiliation{Fermi National Accelerator Laboratory, Batavia, Illinois 60510, USA}
\author{Y. Lu}
\affiliation{Fermi National Accelerator Laboratory, Batavia, Illinois 60510, USA}
\author{G. Eremeev}
\affiliation{Fermi National Accelerator Laboratory, Batavia, Illinois 60510, USA}
\author{A. Romanenko}
\affiliation{Fermi National Accelerator Laboratory, Batavia, Illinois 60510, USA}
\author{A. Grassellino}
\affiliation{Fermi National Accelerator Laboratory, Batavia, Illinois 60510, USA}

\date{\today}
\begin{abstract}

Trapped magnetic vortices in niobium introduce microwave losses that degrade the performance of superconducting resonators. While such losses have been extensively studied above 1~K, we report here their direct quantification in the millikelvin and low-photon regime relevant to quantum devices. Using a high-quality factor 3-D niobium cavity cooled through its superconducting transition in controlled magnetic fields, we isolate vortex-induced losses and find the resistive component of the sensitivity to trapped flux $S$ to be approximately 2~n$\Omega$/mG at 10~mK and 6~GHz. The decay rate is initially dominated by two-level system (TLS) losses from the native niobium pentoxide, with vortex-induced degradation of $T_1$ occurring above $B_{\text{trap}}\sim$~50~mG. In the absence of the oxide, even 10~mG of trapped flux limits performance $Q_0\sim$~10$^{10}$, or $T_1\sim$~350~ms, underscoring the need for stringent magnetic shielding. The resistive sensitivity $S$ decreases with temperature and remains largely field-independent, whereas the reactive component, $S'$, exhibits a maximum near 0.8~K. These behaviors are well modeled within the Coffey-Clem framework in the zero-creep limit, under the assumption that vortex pinning is enhanced by thermally activated processes. Our results suggest that niobium-based transmon qubits can tolerate vortex-induced dissipation at trapped field levels up to several hundred mG, but achieving long coherence times still requires careful magnetic shielding to suppress lower-field losses from other mechanisms.

\end{abstract}

\maketitle

The widespread adoption of niobium for quantum computing applications has enabled \mbox{three-dimensional (3-D)} superconducting radio-frequency (SRF) resonators with lifetimes $T_1$ up to 2~seconds of coherence \cite{Rom20,Rosenblum_PRXQuantum_2023} and two-dimensional qubits with $T_1\sim$~600~$\mu$s \cite{Bal_npjQuantumInfo_2024}. However, further improvements are necessary to realize a quantum computer capable of practical computation. This requires identifying and mitigating every loss channel present in complex multilayer, nonlinear, qubit elements, including quasiparticles \cite{DeVisser_PRL_2014, deGraff_SciAdv_2020}, radiation \cite{Martinis_Nature_2021}, and two-level systems (TLS) \cite{Muller_IOP_2019}. Recent efforts have identified several such sources of decoherence in these integrated systems, including highly-lossy silicon substrates \cite{Checchin_PRApplied_2022} and oxygen vacancies in the niobium pentoxide as a source of TLS \cite{Rom20,Bafia_PRApplied_2024,Pritchard_ArXiv_2024}. On the other hand, high $Q_0$ performance in niobium-based resonators has been demonstrated to be relatively insensitive to material purity at low fields and millikelvin temperatures \cite{Bektur_APL_2024}, highlighting the robustness and practical advantages of niobium in quantum computing applications. These findings have aided in the development of mitigation strategies which yield substantially improved \mbox{two-dimensional (2-D)} transmon qubit metrics \cite{Bal_npjQuantumInfo_2024}. However, there remains a phenomenon of superconducting niobium which has not yet been characterized at the fields and temperatures relevant for \mbox{2-D} and \mbox{3-D} quantum computing architectures: trapped magnetic flux.

The effect of trapped magnetic flux under the influence of radio-frequency (RF) fields in niobium has been widely studied in the context of particle accelerator applications \cite{Padamsee98}. Under ideal conditions, cooling a niobium cavity through the transition temperature $T_c$ is expected to expel incident static magnetic field via the Meissner effect. However, lattice defects and other inhomogeneities may pin flux lines and trap vortices. The vortices then interact with incident RF fields and introduce additional surface resistance via $R_{\text{s}}\approx R_{\text{T}}+R_{\text{0}}+R_{\text{Fl}}$, where $R_\text{{T}}$, $R_{\text{0}}$, and $R_{\text{Fl}}$ are the surface resistance due to quasiparticles, material properties, and trapped magnetic flux, respectively. We note that this decomposition is approximate, as loss mechanisms may not be strictly additive. Moreover, it has been shown that large thermal gradients supply a thermal depinning force which migrates the vortices either to regions of lower RF magnetic field or removes them altogether from the cavity \cite{Romanenko_JAP_2014}. This finding has led to the development of cooldown protocols which enable ultra-high $Q_0>$~2$~\times$~10\textsuperscript{11} post cooling in ambient magnetic fields of 190~mG \cite{Romanenko_APL_2014}. We note, however, that such cooldown protocols are not achievable in the slow and nearly homogeneous cooling provided by dilution refrigerators (DR), indicating that any incident magnetic field is trapped in devices. To quantify these losses, previous studies have introduced the sensitivity to trapped magnetic flux $S = R_{\text{Fl}}/B_{\text{trap}}$, where $B_{\text{trap}}$ is the amount of trapped magnetic field in the cavity walls \cite{Martinello_APL_2016}. Varying the electronic mean free path (MFP) within the penetration depth in niobium cavities results in a bell-shaped dependence of $S$, ranging from 0.4-1.5~n$\Omega$/mG at 5~MV/m at a temperature of 1.5~K and resonant frequency ($f_0$) of 1.3 GHz \cite{Martinello_APL_2016}. This behavior arises from the interplay between the flux-pinning regime, where magnetic vortices are strongly pinned and contribute minimally to dissipation, and the flux-flow regime, where vortices move freely under radio-frequency (RF) currents and dissipate energy more steadily. The maximum dissipation occurs in the intermediate regime, where vortices are partially depinned and their motion is most responsive to the RF drive \cite{Checchin_APL_2018}. This sensitivity to trapped magnetic flux has not been explored in niobium at the low electric fields and millikelvin temperatures relevant for quantum computing and may potentially introduce substantial degradation in complex, multi-layer qubit systems.

In this study, we directly quantify the dissipation introduced by trapped magnetic flux in niobium at millikelvin temperatures and low photon counts using a single interface, high-quality factor niobium SRF cavity. By cooling this cavity in a dilution refrigerator under controlled applied magnetic fields, we deliberately trap various amounts of magnetic vortices in the cavity walls. We determine that the sensitivity to trapped magnetic flux is $S = 2 ~\text{n}\Omega$/mG at $T=$~10~mK and $f_0=$~6~GHz, with sensitivity decreasing with temperature. The vortex-induced frequency shift is used to extract the reactive component of the sensitivity $S'$, which peaks near 0.8~K. These behaviors are well described by assuming the thermal activation of pinning centers within the Coffey-Clem framework, which models the complex resistivity arising from vortex motion under oscillatory driving currents \cite{CC_PRL_1991}. Our results demonstrate that in fully oxidized 6 GHz niobium cavities, non-vortex losses limit $T_1$ to approximately 25~ms, while vortex driven losses begin to significantly impact $T_1$ above $B_{\text{trap}}\sim$~50~mG. In contrast, in the absence of niobium pentoxide, even modest fields of $B_{\text{trap}}\sim$~10~mG will limit 3-D niobium cavities to $T_1\sim$~350~ms.

\begin{figure}
    \centering
    \includegraphics[width=8.5cm]{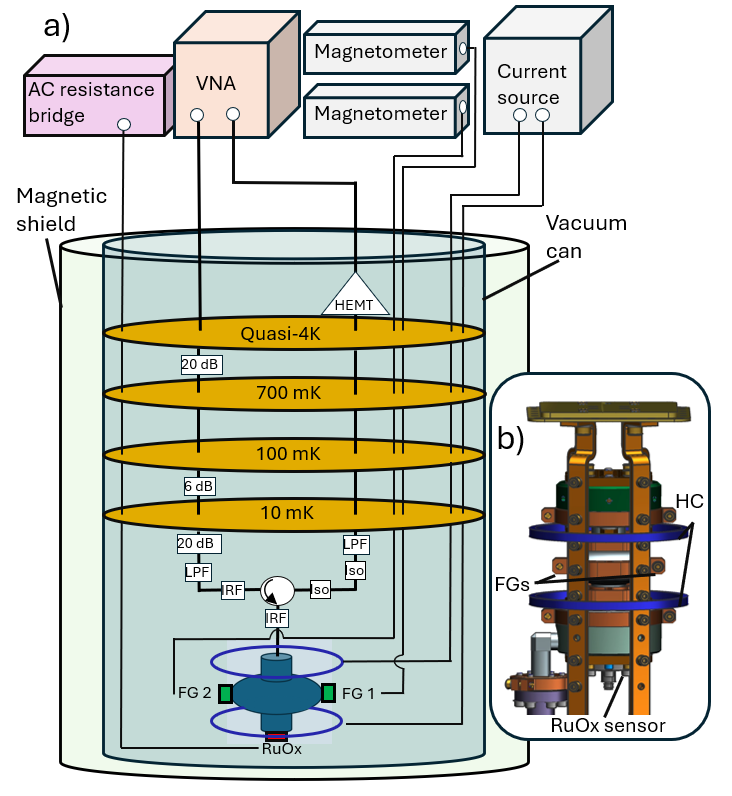}
    \caption{a) Cartoon schematic of the setup used for our study. The measurement chain includes 46~dB of attenuation, low pass filters (LPF), infrared radiation filters (IRF), a circulator, isolators (Iso) and +35 dB HEMT amplifier. A double layer of mu-metal magnetic shielding was used to minimize the ambient magnetic field within the DR. The Helmholtz coils (HC) are shown in blue. Flux gate (FG) and RuOX temperature (RuOx) sensor positions are also presented. b) A depiction of the 6~GHz cavity with thermal anchoring and the precise location of the Helmholtz coils and diagnostic equipment.}
    \label{fig:DRSetup}
\end{figure}

Fig.~\ref{fig:DRSetup}a) illustrates the experimental setup. We employed a 6~GHz niobium TESLA-shaped SRF cavity \cite{Aune00} that underwent a standard treatment process--removing 160~$\mu$m from the inner surface via buffer chemical polishing, followed by annealing at 800$^{\circ}$C for 3 hours in ultra-high vacuum and subsequent high pressure rinsing \cite{Padamsee98}. 

As shown in Fig.~\ref{fig:DRSetup}b), the cavity was thermally anchored to the mixing chamber (MXC) plate of a dilution refrigerator and equipped with two single-axis flux gates at the equator along with a RuOx temperature sensor at the bottom flange. Helmholtz coils were mounted on either side of the cavity cell, facilitating cooling down through the transition temperature of 9.2~K in distinct magnetic field environments, thus varying the level of magnetic vortices trapped in the cavity walls. Table~\ref{tab:magField} presents data on the measured magnetic field just above ($B_{NC}$) and below ($B_{SC}$) the niobium superconducting transition temperature for each of the four cooldowns discussed here. Between each cooldown, we warmed the refrigerator to at least 15~K and adjusted the applied magnetic field. The near-unity ratio of $B_{SC}/B_{NC}$ indicates that nearly all incident flux is trapped within the cavity walls, reflecting inefficient flux expulsion consistent with spatially uniform thermal gradients during cooldown \cite{Posen_JAP_2016}. This suggests that $B_{\text{trap}}\approx B_{NC}$. Moreover, the consistent ratio of 1.02 across cooldowns indicates reproducible thermal conditions from run to run. Once the cavity was well below $T_c$, the Helmholtz coils and flux gates were powered off to minimize heat deposition. The ambient magnetic field near the cavity with the Helmholtz coils off was $-$10~$\pm$~6.1~mG.

We measured the cavity quality factor using methods similar to those described in Ref. \cite{Rom20}. With a vector network analyzer, we first measured the resonance frequency. We then performed time-domain zero span decay measurements of the transmitted power $P_\text{T}(t)$ at this frequency after shutting off the incident signal. The first derivative of the resulting curve was used to obtain field-resolved measurements of the cavity loaded quality factor $Q_{\text{L}}$. We extracted the intrinsic quality factor via $Q_0 = (1/Q_{\text{L}}-1/Q_1)^{-1}$, where $Q_1$~=~1.4~$\times$~10\textsuperscript{9} is the antenna quality factor obtained by circle fitting reflection scattering parameter data. We swept the DR temperature from 10~mK to 1.3~K by applying current to a heater located on the MXC plate. At each intermediate temperature, we allowed the DR to thermalize for at least two hours while acquiring decay data, verified by the RuOx temperature sensor. Thermalized data was averaged and used for our analysis. We used the amplified Johnson-Nyquist thermal noise from the HEMT amplifier to estimate the actual $P_\text{T}$ emitted from the cavity and obtained the on-axis electric field via $E \propto  \sqrt{P_\text{T} Q_1}$. The intracavity photon number was determined using $n=U/\hbar \omega$, where $U$ and $\omega$ are the cavity stored energy and angular frequency.

\begin{table}
    \centering
    \begin{tabular}{cccccc}
        \hline
        CD\#  & Current (mA) & $B_{NC}~(\text{mG})$ & $B_{SC}~(\text{mG})$ & $B_{SC}/B_{NC}$\\
        \hline
        \hline
         1 & 3.45 & 0.0~$\pm$~5.2 &  -0.1~$\pm$~6.4 & -\\
         2 & 20.25 & 50.4~$\pm$~7.5 &  51.3~$\pm$~7.7 & 1.02~$\pm$~0.22\\
         3 & 37.20 & 100.9~$\pm$~9.0 &  102.5~$\pm$~9.3 & 1.02~$\pm$~0.13\\
         4 & 87.46 & 250.5~$\pm$~13.3 &  254.8~$\pm$~13.7 & 1.02~$\pm$~0.08\\
    \end{tabular}
    \caption{Flux gate readings for each cooldown (CD) above ($B_{NC}$ $\approx B_{\text{trap}}$) and below ($B_{SC}$) the niobium transition temperature along with the current applied to the Helmholtz coils. Error values capture the difference in magnetic field readings between the two flux gates due to the presence of magnetic contamination, such as the right-angle valve of the cavity.}
    \label{tab:magField}
\end{table}

Fig.~2 presents a subset of the data acquired from the four cooldowns. For cooldown \#1, in the absence of trapped magnetic flux, we observe the expected behaviors as a function of both temperature and field due to the gradual saturation of TLS localized in the niobium oxide \cite{Bafia_PRApplied_2024}. At low temperatures and fields, TLS contribute strongly to microwave loss, leading to a reduced quality factor; as temperature or field increases, the TLS saturate, resulting in an increase in $Q_0$. As the level of trapped magnetic field in the cavity walls increases, we find an overall reduction in $Q_0$ as a function of temperature and field.

\begin{figure}
    \centering
    \includegraphics[width=8.6cm]{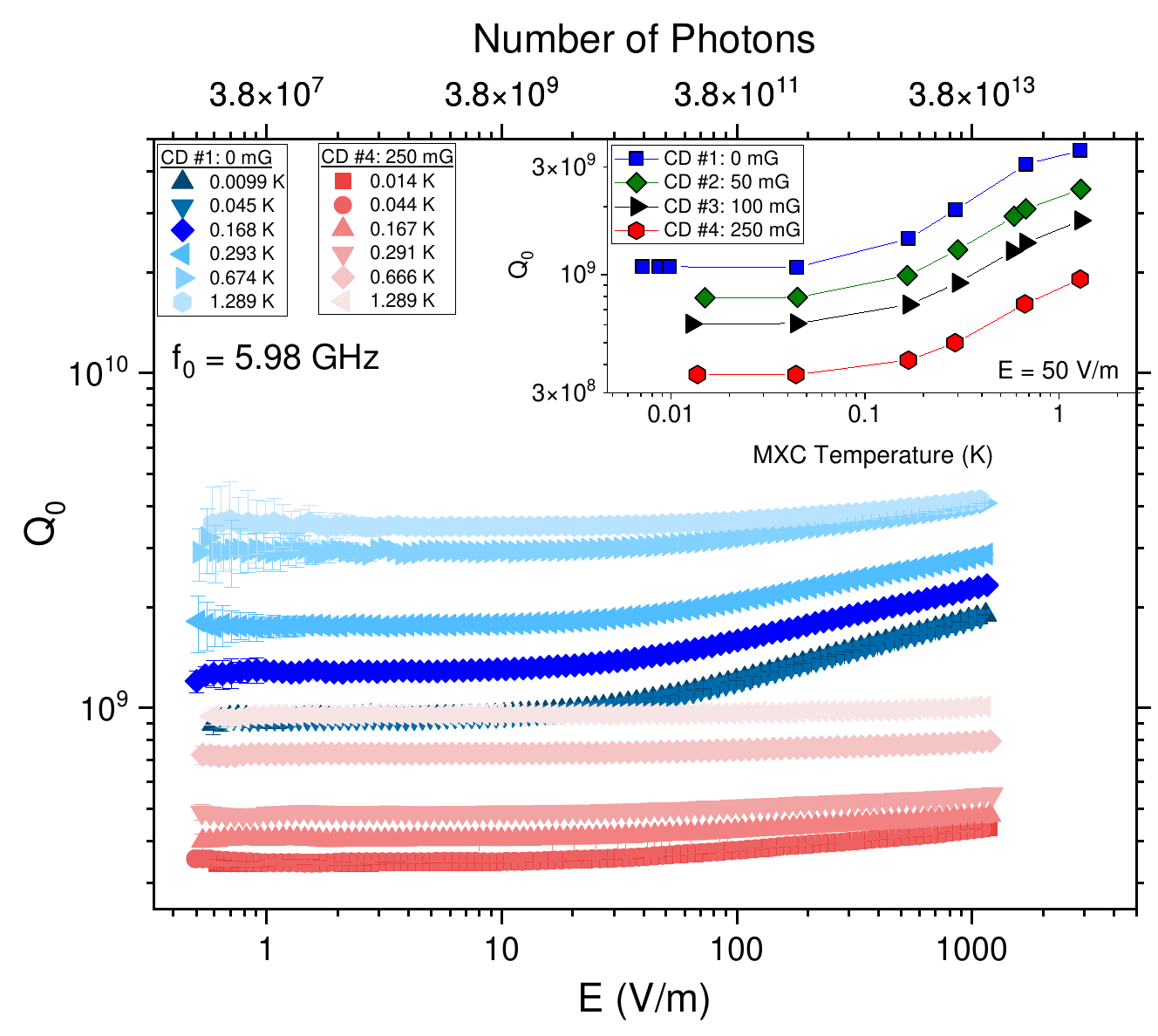}
    \caption{Quality factor vs on-axis electric field of a 6~GHz cavity measured at temperatures ranging from 0.010~K to 1.289~K after cooling in various applied magnetic fields. Inset shows quality factor measured at an on-axis field of 50~V/m vs temperature. Blue, green, black, and red-hued points correspond to data acquired post cooling in 0~mG, 50~mG, 100~mG, and 250~mG, as described in Table~\ref{tab:magField}.}
    \label{fig:Data}
\end{figure}

\begin{figure}
    \centering
    \includegraphics[width=8.6cm]{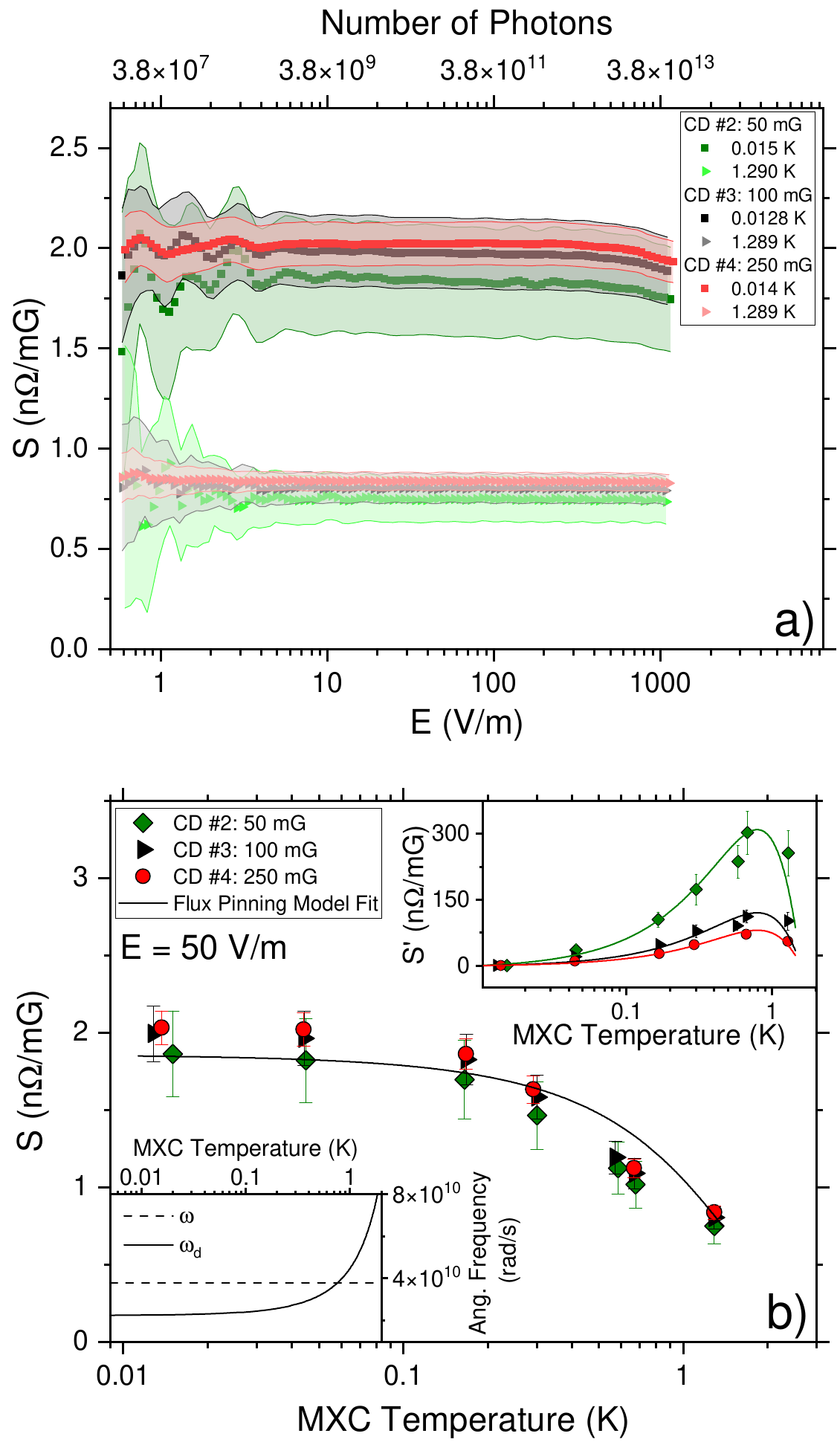}
    \caption{Real part of the sensitivity to trapped magnetic flux as a function of a) field at temperatures of approximately 0.014~K and 1.289~K and b) temperature at 50~V/m post cooling in various magnetic fields. Upper inset plots the imaginary component of the sensitivity to trapped magnetic flux as a function of temperature. Error bars represent the propagated uncertainty arising from data scatter and the applied magnetic field uncertainty. Solid lines in b) represent simultaneous fits to both $S$ and $S'$ for all three data sets using Eq.~\ref{eq:ZT}. Lower inset compares the cavity angular frequency $\omega$ with the fitted depinning frequency $\omega_d$. Fitting parameters are presented in Table~\ref{tab:fit}.}
    \label{fig:Sensitivity}
\end{figure}

We now extract the resistive (real) and reactive (imaginary) components of the sensitivity to trapped magnetic flux, denoted as $S$ and $S'$, respectively. Vortex-induced losses are isolated in CD \#2-4 by subtracting off the contribution from material origins obtained in CD \#1 and normalizing by the level of trapped field, such that
\begin{equation}
    S = \frac{R_{\text{Fl}}}{B_{\text{trap}}}= \frac{G}{B_{\text{trap}}}\bigg(\frac{1}{Q_{0,\text{n}}}-\frac{1}{Q_{0,\text{1}}}\bigg),
\label{eq:S}
\end{equation}

\begin{equation}
    S'= \frac{\Delta X_{\text{Fl}}}{B_{\text{trap}}}= -\frac{2G}{B_{\text{trap}}}\bigg(\frac{f_{0,n}-f_{0,1}}{f_{0,1}(0)}\bigg),
\label{eq:S'}
\end{equation}
where $\Delta X_{\text{Fl}}$ is the vortex-induced shift in surface reactance, $G=$~275~$\Omega$ is a geometric constant, $Q_{0,\text{n}}$, $Q_{0,\text{1}}$, $f_{0,n}$, and $f_{0,1}$ are the intrinsic quality factors and resonant frequencies measured from \mbox{CD~\#n~=~2-4} and \mbox{CD~\#1}, respectively, and $f_{0,1}(0)$ is the resonant frequency from CD \#1 at near-zero temperature. The results are presented in Fig.~\ref{fig:Sensitivity}. The resistive component of the sensitivity data obtained from different cooldown cycles agree within the error bars, reinforcing confidence in the trapped field levels in the cavity walls and demonstrating repeatability of our measurements. In Fig.~\ref{fig:Sensitivity}a), $S$ remains largely independent of $E$ below fields of 1000~V/m. To maintain plot clarity, only results obtained near 0.014~K and 1.289~K are presented, as data collected at intermediate temperatures exhibit similar behaviors. In contrast, Fig.~\ref{fig:Sensitivity}b) reveals that $S$ saturates below approximately 100~mK to a value of 2~n$\Omega$/mG; as temperature increases, sensitivity decreases. Moreover, the reactive component of the trapped magnetic flux exhibits a pronounced maximum near 0.8~K, with lower levels of trapped magnetic field producing larger peak magnitudes. 

We compare our experimental results with those of Martinello \textit{et al.} \cite{Martinello_APL_2016}, who studied similarly processed 1.3~GHz cavities at 1.5~K and in the MV/m field range. To account for differences in experimental conditions, we consider variations in frequency, field amplitude, and temperature. Trapped-flux-induced surface resistance is expected to scale as $R_{\text{Fl}}\sim \sqrt{f_0}$ \cite{Calatroni_PRAB_2019}, and we assume a linear response of fluxoids to the RF field, such that $S$ remains field-independent up to MV/m-level gradients. From Fig.~\ref{fig:Sensitivity}b), we extract $S\approx~$0.7~n$\Omega$/mG at 1.5~K; applying frequency scaling yields a value of approximately 0.32~n$\Omega$/mG. This result is consistent with the values reported by Martinello \textit{et al.} for cavities treated with similar baking and buffer chemical polishing procedures.

A modified Coffey-Clem (CC) model of vortex-driven dissipation under RF excitation is employed to describe the observed temperature dependence of the resistive and reactive responses presented in Fig.~\ref{fig:Sensitivity} \cite{CC_PRL_1991}. In the zero-vortex-creep limit, which is relevant at the millikelvin temperatures explored here, the CC resistivity due to vortex motion reduces to the Gittleman-Rosenblum formalism. This is expressed as $\rho_{\text{GR}}=\rho_{ff}/(1-\text{i}\omega_p/\omega)$, where $\rho_{ff}$ is the flux-flow resistivity, $\omega$ is the angular frequency of the applied RF field, and $\omega _d$ is the depinning frequency, proportional to the vortex pinning constant $k$ \cite{Gittleman_PRL_1966,Alimenti_SUST_2021}. According to Bardeen and Stephen, $\rho _{ff}=\rho _n B_{\text{trap}}/B_{c2}(T)$ \cite{BS_PR_1965}, where $\rho _n$ is the normal-state resistivity and $B_{c2}(T)=B_{c2}(0)(1-(T/T_{c})^2)$ is the temperature-dependent upper critical field. At low temperatures, the complex surface impedance can be approximated as $Z(T)\approx\text{i}\omega \mu _0 (\lambda_s ^2(T) + \lambda_v ^2(T))^{1/2}$, where $\lambda_s(T)$ and $\lambda_v(T)$ are the condensate and vortex penetration depths \cite{Golosovsky_SUST_1996}. Using the relation $\lambda_v =(\text{i} \rho _{ff}/\mu_0 \omega)^{1/2}$ \cite{Golosovsky_SUST_1996} and expanding $Z(T)$ in the small-field limit and normalizing by the trapped magnetic field yields

\begin{equation}
    \frac{Z_{\text{Fl}}(T)}{B_{\text{trap}}}
    = S + \text{i}S'
    = \frac{\rho _n}{2 \lambda_s(T) B_{c2}(T)} 
    \frac{\omega^2 + \text{i} F \omega \omega_d(T)}{\omega^2 + \omega_d^2(T)},
\label{eq:ZT}
\end{equation}

\noindent where $\omega _d (T) = \omega _0 e ^{\alpha T}$ is the thermally activated depinning frequency. The dimensionless scaling parameter $F$ accounts for systematic variations not explicitly captured by our model, such as non-uniform field distribution and geometric factors. We take $\lambda_L=$~39~nm \cite{Maxfield_PhysRev_1965}, $B_{c2}(0)=0.2$~T, and $\rho _n = 4\times10^{-10}~\Omega$~m; when assuming a material constant of $\rho l=6\times10^{-16}~\Omega$~m, \cite{Hasan_book2}, this latter parameter corresponds to an electronic mean free path $l$ of 1500~nm, in agreement with reported values of similarly treated cavities \cite{Koufalis_IPAC18, Martinello_APL_2016}. Model parameters were extracted by simultaneously fitting both the resistive and reactive responses across all three cooldown datasets. In these fits, the parameters $\omega_0$ and $\alpha$ were shared globally across datasets for both $S$ and $S'$, while $F$ was allowed to vary independently for each cooldown. Fitted parameters are summarized in Table~\ref{tab:fit}. The corresponding model fits are shown as solid lines in Fig.~\ref{fig:Sensitivity}b). The lower inset of Fig.~\ref{fig:Sensitivity}b) compares the cavity angular frequency with the fitted thermally activated depinning frequency, which increases with temperature, suggesting an \textit{increase} in the vortex pinning strength. Notably, for $T>$~0.8~K, the condition $\omega<\omega_d$ is satisfied, indicating a crossover from the flux-flow to the pinning regimes. This crossover temperature coincides with the maximum observed in $S'$ in Fig.~\ref{fig:Sensitivity}b).

\begin{table}[h!]
\centering
\begin{tabular}{ccccc}
\hline
\(\omega_0\) & \(\alpha\) & \(F_{\text{CD1}}\) & \(F_{\text{CD2}}\) & \(F_{\text{CD3}}\) \\
\([10^{10}~\text{rad/s}]\) & \([10^{-1}~\text{K}^{-1}]\) & \([10^3]\) & \([10^2]\) & \([10^2]\)\\
\hline
$2.22(11)$ & $7.01(58)$ & $1.91(35)$ & $7.43(13)$ & $4.97(81)$ \\
\hline
\end{tabular}
\caption{Fitting parameters and uncertainties extracted from simultaneous fits to the three $S$ and $S'$ datasets in Fig.~\ref{fig:Sensitivity}(b) using Eq.~\ref{eq:ZT}.}
\label{tab:fit}
\end{table}

The observed increase in depinning frequency with temperature may arise from several mechanisms. One possibility is the formation of normal-conducting hydride phases NbH\textsubscript{x$>$0.7}, which have been recently identified in niobium SRF cavity cutouts \cite{Sung_2024} and are known to exhibit $T_c <$~1.3~K \cite{Welter_Johnen_PhysikB_1977}. As the temperature increases past this threshold, the hydrides become only weakly superconducting via the proximity effect \cite{Rom13}, potentially serving as strong pinning centers and thereby enhancing the overall pinning strength. Alternatively, thermal suppression of vortex tunneling between pinning centers may contribute. As discussed by Ao and Thouless, the tunneling rate at zero temperature decreases exponentially with temperature, leading to reduced vortex mobility \cite{Ao_Thouless_PRL_1994}. This exponential suppression of tunneling would effectively increase the pinning strength and result in an exponentially increasing depinning frequency.

\begin{figure}
    \centering
    \includegraphics[width=8.6cm]{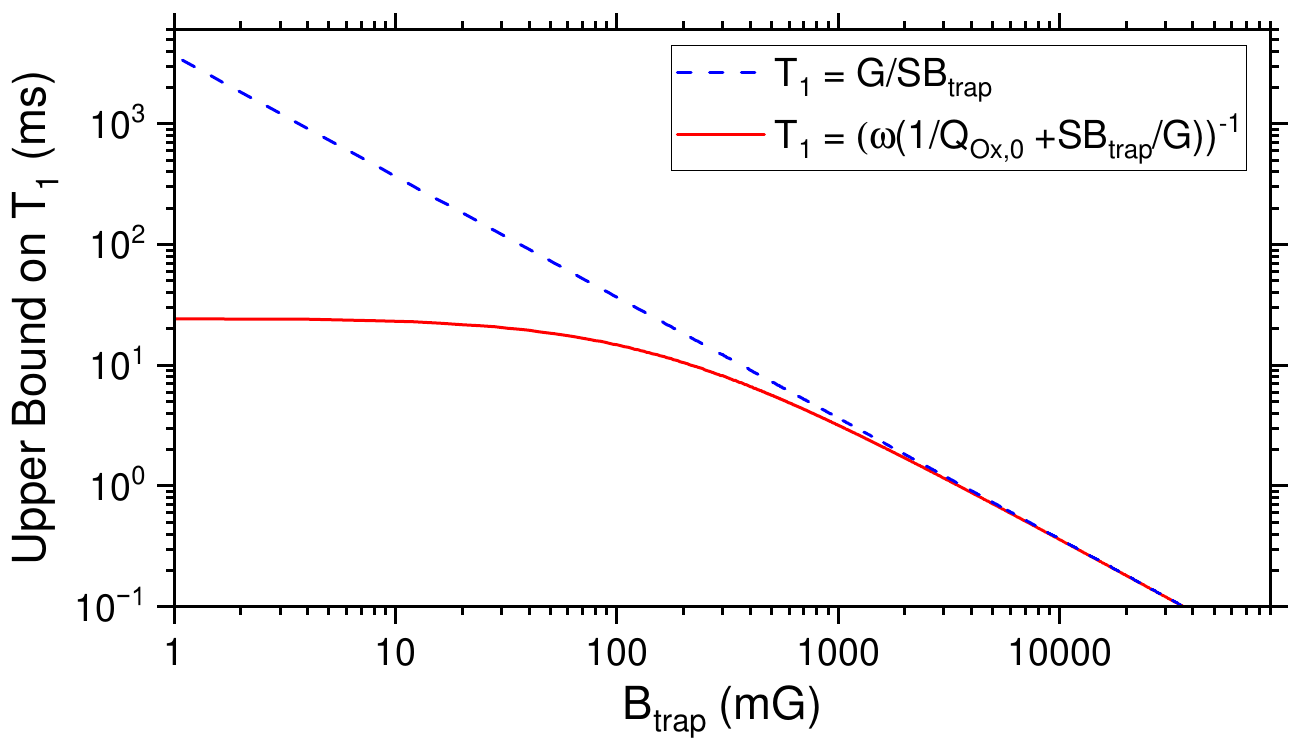}
    \caption{Calculated upper bounds on $T_1$ of a niobium SRF cavity at 6~GHz at 1~V/m considering convolved oxide and vortex contributions (solid red line) and vortex losses in the absence of oxide (blue dashed).}
    \label{fig:T1Calc}
\end{figure}

Our results highlight that minimizing trapped magnetic flux is critical for achieving high $Q_0$ in niobium cavities that are not otherwise dominated by other loss mechanisms, such as those associated with the niobium pentoxide \cite{Rom20,Bafia_PRApplied_2024}. To illustrate this, we plot $T_1 = \omega ^{-1}[(1/Q_{\text{Ox},0})+(S B_{\text{trap}}/G)]^{-1}$ for a 6~GHz niobium SRF cavity, where $Q_{\text{Ox},0}$ is the zero temperature quality factor of a fully oxidized niobium cavity; the result is presented in Fig.~\ref{fig:T1Calc}. For $B_{\text{trap}}<$~25~mG, the native niobium oxide limits $T_1$ to 25~ms; trapped flux begins to significantly impact $T_1$ for $B_{\text{trap}}$ levels of $>$50~mG. In the absence of Nb\textsubscript{2}O\textsubscript{5}, we predict that $B_{\text{trap}}=$~10~mG would yield $T_1\sim$~350~ms, in rough agreement with the 5~GHz cavity post \textit{in situ} oxide removal shown by Romanenko \textit{et al.} \cite{Rom20}.

These results further suggest that transmon qubits which utilize niobium capacitor pads should be robust against trapped magnetic vortex losses on the order of several hundreds of milligauss. This is in agreement with previous studies, where aluminum and rhenium qubits show little additional loss post cooling in applied fields of 500~mG \cite{Song_PRB_2009}. Instead, vortices may be used as a potential strategy for reducing the impact of quasiparticles in qubits \cite{Wang_Nature_2014}. Present studies are underway to quantify these phenomena in niobium pad transmon qubits. Nevertheless, we emphasize the importance of magnetic shielding and maintaining magnetic hygiene to minimize these losses.

In conclusion, we have shown that trapped magnetic vortices can drive a non-negligible level of RF loss in niobium at millikelvin and at low electric fields, relevant for 3-D resonators and 2-D devices. We find that the resistive and reactive components of vortex driven dissipation are well described by a modified Coffey-Clem model, wherein pinning strength increases with temperature. At 10~mK and 6~GHz, trapped flux contributes losses of 2~n$\Omega$/mG. Consequently, in the absence of niobium pentoxide, trapping 10~mG would limit 3-D cavity photon lifetimes to $T_1\sim$~350~ms. We hypothesize that trapped magnetic flux has a relatively small impact on the niobium capacitor pads of transmon qubits for fields less than several hundreds of milligauss; we are directly investigating these losses in a separate experimental study. Our results underscore the crucial role of magnetic shielding and magnetic hygiene in minimizing flux-related losses, required to enable quantum computing and sensing architectures with ultra-long coherence lifetimes of $T_{\text{1}}>$~1~s.

We would like to thank James Sauls, Mehdi Zarea, and Andrei Lunin for insightful discussions. This material is based upon work supported by the U.S. Department of Energy, Office of Science, National Quantum Information Science Research Centers, Superconducting Quantum Materials and Systems Center (SQMS) under Contract No. DE-AC02-07CH11359.

The data that support the findings of this study are available from the corresponding author upon reasonable request.

\bibliography{Main}

\end{document}